\documentclass[twocolumn,prl]{revtex4}
\usepackage{graphicx}
\newcommand{\be}{\begin{equation}}
\newcommand{\ee}{\end{equation}}
\newcommand{\bea}{\begin{eqnarray}}
\newcommand{\eea}{\end{eqnarray}}

\newcommand{\lp}{\left(}
\newcommand{\rp}{\right)}

\renewcommand{\phi}{\varphi}
\renewcommand{\epsilon}{\varepsilon}
\renewcommand{\vec}[1]{{\bf #1}}

\begin{document}

\title{Dissipative Quantum Hall Effect in Graphene near the Dirac Point}
\author{
Dmitry A. Abanin,${}^1$ Kostya S. Novoselov,${}^2$ Uli Zeitler,${}^3$  Patrick A. Lee,${}^1$  Andre K. Geim,${}^2$\footnote{emails: levitov@mit.edu, geim@manchester.ac.uk}
 Leonid S. Levitov${}^{1\ast}$ 
}
\affiliation{
 ${}^1$ Department of Physics,
 Massachusetts Institute of Technology, 77 Massachusetts Ave,
 Cambridge, MA 02139\\
${}^2$ Department of Physics and Astronomy, University of Manchester, 
Manchester, M13 9PL, UK\\
${}^3$ High Field Magnet Laboratory, IMM, Radboud University Nijmegen,  
6525 ED Nijmegen, The Netherlands
}

\begin{abstract}
We report on the unusual nature of $\nu = 0$ state 
in the integer quantum Hall effect (QHE) in graphene 
and show that electron transport in this regime is dominated 
by counter-propagating edge states.
Such states, intrinsic to massless Dirac quasiparticles,
manifest themselves in a large longitudinal resistivity $\rho_{xx}\gtrsim h/e^2$, 
in striking contrast to $\rho_{xx}$ behavior in the standard QHE.
The $\nu=0$ state in graphene is also predicted to exhibit 
pronounced fluctuations in $\rho_{xy}$ and $\rho_{xx}$ 
and a smeared zero Hall plateau in $\sigma_{xy}$, 
in agreement with experiment. The existence of gapless edge
states puts stringent constraints on possible
theoretical models of the $\nu=0$ state.
\end{abstract}

\maketitle
Electronic properties of graphene has attracted significant interest, 
especially since an anomalous integer quantum Hall effect (QHE)
was found in this material\,\cite{Novoselov05,Zhang05}. 
Graphene features QHE plateaus at half-integer values of Hall conductivity 
$\sigma_{xy}=(\pm1/2,\pm3/2,...)4e^2/h$ 
where 
the factor 4 takes into account 
double valley and double spin degeneracy. 
The ``half-integer" QHE is now well understood as arising 
due to unusual charge carriers in graphene, which mimic massless relativistic 
Dirac particles\,\cite{Novoselov04}.
Recent theoretical efforts have focused on the properties of
spin- and valley-split QHE 
at low filling factors\,\cite{Abanin06a,Fertig06,Nomura06,Alicea06,Goerbig06,Abanin06b} and
fractional QHE\,\cite{Apalkov06}. 
Novel states 
with dynamically generated exciton-like gap 
were conjectured near the Dirac point\,\cite{Gusynin94,Khveshchenko01,Gusynin06,Fuchs06}.
Experiments
in ultra-high magnetic fields\,\cite{Zhang06} have so far revealed 
only additional integer plateaus at $\nu =0$, $\pm1$ and $\pm4$, 
which were attributed to valley and spin splitting.

The most intriguing QHE state is 
arguably that observed at $\nu=0$. 
Being intrinsically particle-hole symmetric, 
it has no analog in semiconductor-based QHE systems.
Interestingly, while it exhibits
a step-like feature in $\sigma_{xy}$, the experimentally measured 
longitudinal and Hall resistance\,\cite{Zhang06}
($\rho_{xx}$ and $\rho_{xy}$) 
display neither a clear
quantized plateau nor a zero-resistance state, the hallmarks 
of the conventional QHE. This unusual behavior was attributed 
to sample inhomogeneity\,\cite{Zhang06} and remains unexplained. 
In this Letter, 
we show that such behavior near the Dirac point 
is in fact intrinsic 
to Dirac fermions in graphene and indicates an opening of a spin gap in 
the energy spectrum\,\cite{Abanin06a}. 
The gap leads to counter-circulating edge states carrying opposite spin\,\cite{Abanin06a,Fertig06}
which result in interesting and rather bizarre properties of this QHE state.
In particular, even in the complete absence of bulk conductivity, 
this state has a nonzero $\rho_{xx}\gtrsim h/2e^2$ 
(i.e. the QHE state is dissipative) whereby $\rho_{xy}$ can
change its sign 
as a function of density without exhibiting a plateau. 

\begin{figure}
\includegraphics[width=3.4in]{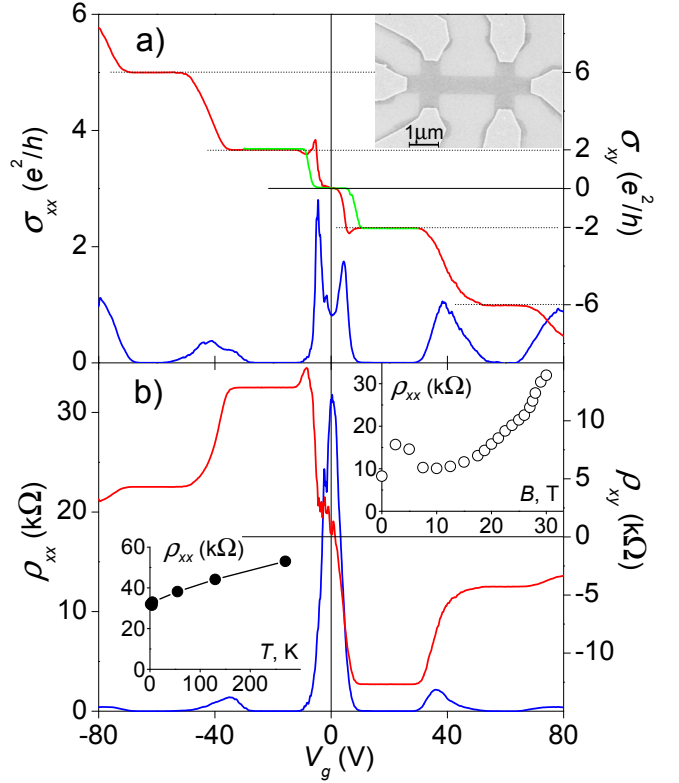}
\vspace{-0.5cm}
\caption[]{
Longitudinal and Hall conductivities $\sigma_{xx}$ and $\sigma_{xy}$ (a)
calculated from $\rho_{xx}$ and $\rho_{xy}$ measured at $4\,{\rm K}$ and $B=30\,{\rm T}$ (b).
The $\nu=0$ plateau in $\sigma_{xy}$ and the double-peak structure in $\sigma_{xx}$ 
arise mostly from strong density dependence of  $\rho_{xx}$ peak
(green trace shows $\sigma_{xy}$ for another sample).
The upper inset shows one of our devices.
Temperature and magnetic field dependence of $\rho_{xx}$
near $\nu=0$ are shown in the insets below.
Note the metal-like temperature dependence of $\rho_{xx}$.
}
\label{fig1}
\end{figure}

We start with reviewing the experimental situation near $\nu =0$. 
Our graphene devices were fabricated as described in Ref.\cite{Novoselov04} 
and fully characterized in fields $B$ up to $12{\rm T}$ at temperatures $T$ down to $1\,{\rm K}$. 
These measurements revealed the behavior characteristic 
of single-layer graphene\,\cite{Novoselov05}. 
Several devices were then investigated 
in $B$ up to $30\,{\rm T}$, 
where, besides 
the standard half-integer QHE sequence, the $\nu =0$ plateau becomes 
clearly visible as an additional step in $\sigma_{xy}$ (Fig.\ref{fig1}). We note, however, 
that the step is not completely flat, and 
there is no clear zero-resistance plateau in $\rho_{xy}$.
Instead, 
$\rho_{xy}$ exhibits a fluctuating feature away from zero 
which seems trying to develop in a plateau (Fig.\ref{fig1}b). 
[In some devices $\rho_{xy}$ passed through zero in a smooth way
without fluctuations.] 
Moreover, $\rho_{xx}$ does not exhibit a zero-resistance state either. 
Instead, it has a pronounced peak  near zero $\nu$ which 
does not split in any field. 
The value at the peak
grows from $\rho_{xx}\approx h/4e^2$ in zero $B$ 
($7.5\,{\rm k}\Omega$ for the shown devices) [1] to 
$\rho_{xx}>45\,{\rm k}\Omega$ at $30\,{\rm T}$ (see inset of Fig.\ref{fig1}b).

At this point, the absence of both hallmarks of the conventional QHE 
in these experiments can make one skeptical 
about the relation between the observed 
extra step in $\sigma_{xy}$ and an additional QHE plateau. 
However, the described high-field behavior near $\nu=0$ was found to be universal 
(reproducible for different samples, measurement geometries 
and magnetic fields above $20\,{\rm T}$). 
It is also in agreement 
with that reported in Ref.\cite{Zhang06}. 
Moreover, one can generally argue that the QHE at $\nu=0$
cannot possibly exhibit the usual hallmarks. Indeed, $\rho_{xy}$ 
has to pass through zero because of the carrier-type change 
but $\rho_{xx}$ cannot simultaneously exhibit a zero-resistance 
state because zero in both $\rho_{xy}$ and $\rho_{xx}$ would indicate 
a dissipationless (superconducting) state. 

\begin{figure}
\includegraphics[width=3.1in]{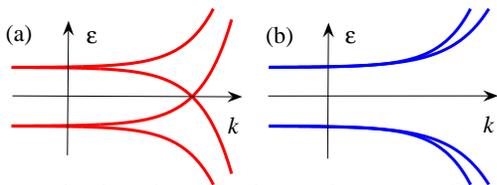}
\vspace{-0.5cm}
\caption[]{
Excitation dispersion in $\nu=0$ graphene QH state for a system with 
and without gapless
chiral edge modes, (a) and (b) respectively.
Case (a) is realized in 
spin-polarized $\nu=0$ state\,\cite{Abanin06a},
while case (b) occurs when symmetry is incompatible
with gapless modes, for example, in valley-polarized $\nu=0$
state conjectured in Ref.\cite{Zhang06}. In the latter a gap
opens at branch crossing due to valley mixing at the 
sample boundary.
}
\label{fig2}
\end{figure}

To explain the anomalous behavior of the high-field QHE (Fig.\ref{fig1}), 
we note that all microscopic models near the Dirac point 
can be broadly classified in two groups, QH metal and QH insulator, 
as illustrated in Fig.\ref{fig2}. 
Transport properties in these two cases
are very different. The QH insulator (Fig.\ref{fig2}b) is characterized by 
strongly temperature dependent resistivity
diverging at low $T$. The metallic $T$-dependence observed at $\nu=0$
clearly rules out this scenario.
In the QH metal (Fig.\ref{fig2}a), 
a pair of gapless edge excitations (Fig.\ref{fig2}a) 
provides dominant contribution 
to $\sigma_{xx}$, while transport in the bulk is suppressed by
an energy gap. 
Such {\it dissipative QHE state}
will have 
$\sigma_{xx}\sim e^2/h \gg \sigma_{xy}$, i.e.
nominally small Hall angle and apparently no QHE.
the roles of bulk and edge transport  
here effectively interchange:
The longitudinal response is due to edge states,
while the transverse response is determined mainly by
the bulk properties.

From a general symmetry viewpoint advanced by Fu, Kane and Mele\,\cite{Liang06}
the existence of counter-circulating gapless 
excitations is controled by $Z_2$ invariants, protecting
the spectrum from gap opening at branch crossing.
In the spin-polarized QHE state\,\cite{Abanin06a}
this invariant is given by $\sigma_z$. While for other $\nu =0$ QHE states\,\cite{Gusynin94,Khveshchenko01,Gusynin06,Fuchs06}
such invariants are not known, any viable theoretical model
must present a mechanism to generate gapless edge states.

The metallic temperature dependence indicates strong dephasing 
that prevents onset of localization.
To account for this observation,
we suppose that the mean free path
along the edge is sufficiently large, such that local equilibrium in
the energy distribution is reached
in between backscattering events.
For that, the rate of inelastic processes 
must exceed the elastic backscattering rate:
$\nu_{\rm inel}\gg\nu_{\rm el}$. This situation
occurs naturally in the Zeeman-split QHE state\,\cite{Abanin06a}, since 
backscattering between chiral modes carrying opposite spins
is controlled by spin-orbital coupling 
which is small in graphene.

In the dephased regime, 
the chiral channels are described 
by local chemical potentials, $\phi_{1,2}(x)$, 
whose deviation from equilibrium is related to currents:
\be\label{eq:I12}
I_1=\frac{e^2}{h}\phi_1
,\quad
I_2=\frac{e^2}{h}\phi_2
,\quad
I=I_1-I_2
,
\ee
where $I$ is the total current {\it on one edge}.
In the absence of backscattering
between the channels the currents $I_{1,2}$ are conserved.
In this case, since the potentials $\phi_{1,2}$ are constant along the edge,
transport is locally nondissipative, similar to the usual QHE
\cite{Halperin}.

\begin{figure}
\includegraphics[width=3.5in]{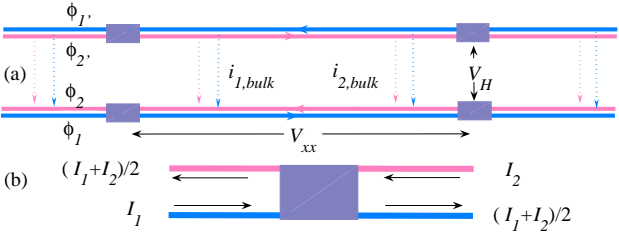}
\vspace{-0.85cm}
\caption[]{
(a) Transport in a Hall bar geometry, Eqs.(\ref{eq:edges_12}).
The edge states denoted by red and blue carry opposite spin.
Transport through the bulk is indicated by dotted lines.
Voltage probes used to measure longitudinal and Hall voltage are shown.
(b) Voltage probe in a full spin mixing regime\,\cite{Buttiker88}
measures $V_{\rm probe}=\frac{h}{2e^2}(I_1+I_2)$.
Note finite voltage drop
across the probe, Eq.(\ref{eq:contact}).
}
\label{fig3}
\end{figure}

The origin of longitudinal resistance in this ideal case
can be traced to the behavior in the contact regions. 
[Note the resemblance of each edge in Fig.\ref{fig3}a 
with two-probe measurement geometry for the standard QHE.]
We adopt the model of termal reservoirs\,\cite{Buttiker88} which assumes full mixing of electron spin 
states within Ohmic contacts (see Fig.\ref{fig3}b). 
With currents $I_1$, $I_2$ flowing into the contact,
and equal currents $I_{1,2}^{\rm (out)}=\frac12(I_1+I_2)$ flowing out, 
the potential of the probe is $V_{\rm probe}=\frac{h}{e^2}I_{1,2}^{\rm (out)}$.
Crucially, using Eq.(\ref{eq:I12}), there is a potential drop across the contact,
\be\label{eq:contact}
\Delta\phi=\frac{h}{2e^2}(I_1-I_2),
\ee
equally for $\phi_1$ and $\phi_2$.
The voltage between
two contacts positioned at the same edge (see Fig.\ref{fig3}a)
is equal to $V_{xx}=\frac{h}{e^2}I$, which gives a universal resistance
value\,\cite{Abanin06a}.
This is in contrast with the usual QHE where there is no voltage 
drop between adjacent potential probes\,\cite{Halperin,Buttiker88}.

The longitudinal resistance increases and becomes nonuniversal 
in the presence of backscattering. 
It can be described by transport equations for charge density 
\begin{eqnarray}\label{eq:edges_12}
&& \partial_t n_1 + \partial_x \phi_1 = \gamma (\phi_2-\phi_1) ,
\cr
&& \partial_t n_2 - \partial_x \phi_2 = \gamma (\phi_1-\phi_2) ,
\quad
n_i=\nu_i\phi_i
,
\end{eqnarray}
where 
$\gamma^{-1}$ is the mean free path for 
1d backscattering between modes $1$ and $2$,
and $\nu_{1,2}$ are compressibilities
of the modes 1 and 2.
In a stationary state, Eqs.(\ref{eq:edges_12}) have an integral
$\tilde I=\phi_1-\phi_2$ which expresses conservation of current
$I=\frac{e^2}{h}\tilde I$. 
The general solution in the stationary current-carrying state
is $\phi_{1,2}(x)=\phi_{1,2}^{(0)}-\gamma x\tilde I$.

For the Hall bar geometry shown in Fig.\ref{fig3}a, taking into account
the contribution of voltage drop
across contacts, Eq.(\ref{eq:contact}),
we find the voltage along the edge $V_{xx}=(\gamma L+1)\tilde I$,
where $L$ is the distance between the contacts. 
In the absence of transport through the bulk, if both edges carry
the same current,   
the longitudinal resistance is
\be\label{eq:edge_Rxx}
R_{xx}=\lp \gamma L+1\rp\frac{h}{2e^2},\quad \rho_{xx}=(w/L)R_{xx},
\ee
with $w/L$ the aspect ratio.
From $\rho_{xx}$ peak value 
(Fig.\ref{fig1}) we estimate 
$\gamma w\approx 2.5$, which 
gives the backscattering mean free path   
of $0.4\,{\rm \mu m}$.
The metallic $T$-dependence of $\rho_{xx}$ signals an increase of scattering
with $T$ (Fig.\ref{fig1}b inset). 
Similarly, $\rho_{xx}$ growing with $B$
is explained by 
enhancement in scattering 
due to electron wavefunction
pushed at high $B$ towards the disordered boundary.

An important consequence of the 1d edge transport is
the enhancement 
of fluctuations caused by position dependence of the scattering rate
$\gamma(x)$. Solving for the potentials at the edge,
\be\label{eq:phi_fluctuations}
\phi_{1,2}(x)=\phi_{1,2}^{(0)}-\tilde I\int_0^x\gamma(y) dy
,
\ee
we see that the fluctuations in the longitudinal 
resistance scale as a square-root of separation between the contacts:
\[
\delta V_{xx}=\tilde I\int_{x_1}^{x_2}\delta \gamma(y)dy
,\quad
\delta R_{xx}\sim (h/e^2)\sqrt{L/a}
\]
where $a\gtrsim\gamma^{-1}$ is a microscopic 
parameter which depends on the
details of spatial correlation of $\gamma(x)$.
Similar effect leads to fluctuations of the Hall voltage
which has zero average value at $\nu=0$. 
Assuming that the fluctuations of the potential at each edge, described
by Eq.(\ref{eq:phi_fluctuations}), are independent, we estimate
$\delta R_{xy}\sim (h/e^2)\sqrt{L/a}$, where $L$ is the bar length.

These fluctuations manifest themselves in noisy features 
in the transport coefficients near $\nu=0$, arising from the dependence of 
the effective scattering potential on electron density.
Such features can indeed be seen in $\rho_{xy}$ 
and $\rho_{xx}$ around $\nu=0$ (Fig.\ref{fig1}b).
As discussed below, away from $\nu=0$ bulk transport becomes important 
and short-circuits the edge.
This will lead to suppression
of fluctuations in $\rho_{xx}$ and $\rho_{xy}$ away from $\nu=0$,
in agreement with the behavior of the fluctuations in Fig.\ref{fig1}b.

Another source of asymmetry in voltage distribution 
on opposite sides of the Hall bar is the potential drop on a contact,
Eq.\,(\ref{eq:contact}). 
This quantity can be nonuniversal for imperfect contacts,
leading to finite
transverse voltage. Such an effect
can be seen in $\rho_{xy}$ data in Fig.\ref{fig1}
near $\nu=0$, 
where Hall effect in a pristine system would vanish. 

To describe transport properties at finite densities around 
$\nu=0$, we must account for transport in the bulk. 
This can be achieved by incorporating in Eq.(\ref{eq:edges_12})
the terms describing the edge-to-bulk leakage:
\begin{eqnarray}\label{eq:phi,psi}
\partial_x \phi_1 = \gamma (\phi_2-\phi_1) 
+g(\psi_1-\phi_1),
\cr
- \partial_x \phi_2 = \gamma (\phi_1-\phi_2) 
+g(\psi_2-\phi_2),
\end{eqnarray}
where $\psi_{1,2}$ are the up- and down-spin electrochemical potentials 
in the bulk near the boundary. Transport in the interior
of the bar is described by tensor current-field relations
with the longitudinal and Hall conductivities
$\sigma^{(1,2)}_{xx}$, $\sigma^{(1,2)}_{xy}$ for each spin component.
Combined with current continuity, these relations yield
the 2d Laplace's equation for the quantities $\psi_{1,2}$,
with boundary conditions supplied by current continuity
at the boundary:
\be\label{eq:bound_conds}
\sigma^{(i)}_{xx}\vec n .\nabla\psi_i
+\sigma^{(i)}_{xy}\vec n \times\nabla\psi_i + g(\phi_i-\psi_i)
=0
,\quad i=1,2,
\ee
where $\vec n$ is a unit normal vector.
[In Eq.(\ref{eq:bound_conds}) and below we use the units of $e^2/h=1$.]
To describe dc current, we seek a solution of Eqs.(\ref{eq:phi,psi})
on both edges of the bar
with linear $x$ dependence
$\phi_i(x)=\phi_i^{(0)}-{\cal E} x$ which satisfies boundary conditions
(\ref{eq:bound_conds}), where the functions $\psi_{1,2}$ have 
a similar linear dependence. 
The current is calculated from this solution
as a sum of the contributions from the bulk and both edges.
After elementary but somewhat tedious algebra we obtain 
a relation $I=2{\cal E}/\tilde\gamma$, where
\be\label{eq:G_total}
\frac2{\tilde\gamma}
=\frac4{2\gamma+g}+ \frac{w}{\rho^{(1)}_{xx}}+\frac{w}{\rho^{(2)}_{xx}}
-\frac{\lambda w\lp \tilde\sigma^{(1)}_{xy}/\sigma^{(1)}_{xx}-\tilde\sigma^{(2)}_{xy}/\sigma^{(2)}_{xx}\rp^2}{2+\lambda/\sigma^{(1)}_{xx}+\lambda/\sigma^{(2)}_{xx}},
\ee
with $w$ the bar width and $\lambda=w g\gamma /(2\gamma+g)$.
The quantities 
$\tilde\sigma^{(1,2)}_{xy}=\sigma^{(1,2)}_{xy}\pm g/(2\gamma+g)$ represent the sum 
of the bulk and edge contributions to Hall conductivities,
and $\rho^{(1,2)}_{xx}$ are defined as
$\rho^{(i)}_{xx}=\sigma^{(i)}_{xx}/(\tilde\sigma^{(i)}_{xy}{}^2+\sigma^{(i)}_{xx}{}^2)$.
The quantity $\tilde\gamma$, Eq.(\ref{eq:G_total}), replaces $\gamma$ in Eq.(\ref{eq:edge_Rxx}).
At vanishing bulk conductivity, $\sigma^{(1,2)}_{xx}\to 0$,
we recover $\tilde\gamma=\gamma$.

\begin{figure}
\includegraphics[width=3.3in]{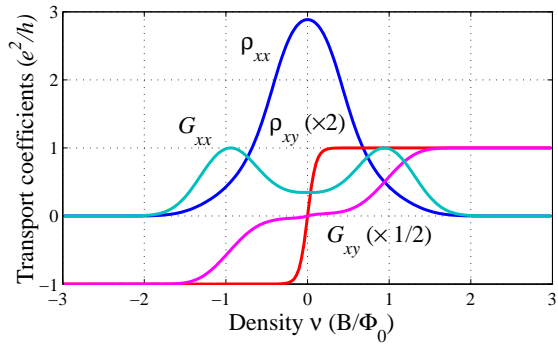}
\vspace{-0.5cm}
\caption[]{
Density dependence of transport coefficients $\rho_{xx}$, $\rho_{xy}$
and $G_{xx}=\rho_{xy}/(\rho_{xy}{}^2+\rho_{xx}{}^2)$, 
$G_{xy}=\rho_{xy}/(\rho_{xy}{}^2+\rho_{xx}{}^2)$
for a Hall bar (Fig.\ref{fig3}), obtained 
from the 
edge transport model (\ref{eq:phi,psi}) augmented with
bulk conductivity (Eqs.(\ref{eq:G,G_H}),(\ref{eq:G_total}),(\ref{eq:G_H}), see text). 
The peak in $\rho_{xx}$ at $\nu=0$ is due to 
the edge contribution,  
shunted by the bulk conductivity away from $\nu=0$. 
Note the smooth behavior of $\rho_{xy}$ near $\nu=0$,
a tilted plateau in  
$G_{xy}$, and a double-peak structure in $G_{xx}$.
}
\label{fig4}
\end{figure}

The Hall voltage can be calculated from this solution as 
$V_{H}=\frac12(\phi_1+\phi_2-\phi_{1'}-\phi_{2'})$, where $\phi_{i,i'}$
are variables at opposite edges. We obtain $V_{H}=\xi{\cal E}$, where
\be\label{eq:G_H}
\xi=2w\frac{\tilde\sigma^{(1)}_{xy}\lp \lambda+\sigma^{(2)}_{xx}\rp
+\tilde\sigma^{(2)}_{xy}\lp \lambda+\sigma^{(1)}_{xx}\rp}{2\sigma^{(1)}_{xx}\sigma^{(2)}_{xx}+\lambda\sigma^{(2)}_{xx}+\lambda\sigma^{(1)}_{xx}}
.
\ee
This quantity vanishes at $\nu=0$, since 
$\sigma^{(1)}_{xy}=-\sigma^{(2)}_{xy}$ and 
$\sigma^{(1)}_{xx}=\sigma^{(2)}_{xx}$ at this point
due to particle-hole symmetry.

In Fig.\ref{fig4} we illustrate 
the behavior of the longitudinal and transverse resistance,
calculated from Eqs.(\ref{eq:G_total}),(\ref{eq:G_H}) as
\be\label{eq:G,G_H}
\rho_{xx}=w\tilde\gamma/2,\quad
\rho_{xy}=\xi\tilde\gamma/2
\ee
with $\gamma w=6$, $g w=1$ (the omitted contact term (\ref{eq:contact}) is small 
for these parameters).
Conductivities $\sigma_{xx}^{(1,2)}$, $\sigma_{xy}^{(1,2)}$
are microscopic quantities, and their detailed 
dependence on the filling factor is beyond the scope of this paper.
Here we model the 
conductivities $\sigma^{(1,2)}_{xx}$
by gaussians centered at $\nu=\pm1$,
$\sigma^{(1,2)}_{xx}(\nu)=e^{-A(\nu\pm1)^2}$,
as appropriate for valley-degenerate Landau level,
whereby $\sigma^{(1,2)}_{xy}$ is related to  $\sigma^{(1,2)}_{xx}$
by the semicircle relation\,\cite{semicircle_relation}:
$\sigma^{(1,2)}_{xy}(\sigma^{(1,2)}_{xy}\mp 2)+(\sigma^{(1,2)}_{xx})^2=0$.
In Fig.\ref{fig4} we used $A=5$, however we note that 
none of the qualitative features depend on the details of the
model.

Fig.\ref{fig4} reproduces many of the key features of the data shown in
Fig.\ref{fig1}. The large peak in $\rho_{xx}$ is due to edge transport
near $\nu=0$. The peak is reduced at finite $\nu$ because the edge is
short-circuited by the bulk conductivity.
  The latter corresponds to the double peak structure in $G_{xx}$ in
Fig.\ref{fig4}. We note that the part of 
$G_{xx}$ between the peaks exceeds the superposition of two Gaussians which
represent the bulk conductivity in our model. This excess in $G_{xx}$ is the
signature of the edge contribution. 
   The transverse resistance $\rho_{xy}$ is nonzero due to imbalance in
$\sigma_{xy}^{(1,2)}$ for opposite spin polarizations away from 
the
particle-hole symmetry point 
$\nu=0$. 
Notably, $\rho_{xy}$ does
not show any plateau in the theoretical curve (Fig.\ref{fig4}), while
$G_{xy}$ calculated from $\rho_{xy}$ and $\rho_{xx}$ exhibits a
plateau-like feature. This behavior is in agreement with experiment
(Fig.\ref{fig1} and Ref.\cite{Zhang06}).

To conclude, QH transport in graphene at $\nu=0$ is due to
counter-circulating edge states. In this dissipative QHE
the roles of the bulk and the edge interchange: 
the edge states dominate in the longitudinal conductance, 
while the bulk conductivity
determines the Hall effect. This model explains
the observed behavior of transport coefficients,
in particular the peak in $\rho_{xx}$ and its field and temperature dependence,
lending strong support to the
chiral spin-polarized edge picture of the $\nu=0$ state.

This work is supported by EPSRC (UK), NSF MRSEC Program (DMR 02132802),
NSF-NIRT DMR-0304019 (DA, LL), and NSF grant DMR-0517222 (PAL).


\end{document}